\documentclass[12pt]{article}
\pdfoutput=1
\usepackage{putex}
\usepackage{feyn}
\usepackage[vcentermath]{youngtab}
\usepackage{subfig}
\usepackage{lscape}

\usepackage{graphicx}
\usepackage{epstopdf}
\usepackage{enumerate}
\usepackage{cite}
\usepackage{tensor}
\usepackage{slashed}
\usepackage{amsmath}
\usepackage{amssymb}
\usepackage{mathrsfs}
\usepackage{lgrind}

\usepackage{hyperref}

\numberwithin{equation}{section}

\newcommand{\OO}{\mathbb{O}}

\newcommand {\be} {\begin {equation}}
\newcommand {\ee} {\end {equation}}

\newcommand {\bes} {\begin {equation*}}
\newcommand {\ees} {\end {equation*}}

\newcommand{\Par}{\partial_{\mu}}

\newcommand{\eps}{\epsilon}

\def\CH{{\cal H}}

\newcommand{\beq}{\begin{equation}}
\newcommand{\eeq}{\end{equation}}

\def\be{ \begin{equation} }
\def\ee{ \end{equation} }

\begin{document}

\preprint{PUPT-2474}

\institution{PU}{Department of Physics, Princeton University, Princeton, NJ 08544}
\institution{PCTS}{Princeton Center for Theoretical Science, Princeton University, Princeton, NJ 08544}

\title{
Three Loop Analysis of the Critical $O(N)$ Models in $6-\eps$ Dimensions
}

\authors{Lin Fei,\worksat{\PU} Simone Giombi,\worksat{\PU} Igor R.~Klebanov\worksat{\PU,\PCTS} and Grigory Tarnopolsky\worksat{\PU}
}

\abstract{We continue the study, initiated in arXiv:1404.1094, of the $O(N)$ symmetric theory of $N+1$ massless scalar fields in $6-\epsilon$ dimensions.
This theory has cubic interaction terms $\frac{1}{2}g_1 \sigma (\phi^i)^2 + \frac{1}{6}g_2 \sigma^3$. We calculate the 3-loop beta functions for the two couplings and use them
to determine certain operator scaling dimensions at the IR stable fixed point up to order $\epsilon^3$. We also use the beta functions to determine the corrections
to the critical value of $N$ below which there is no fixed point at real couplings.  The result
suggests a very significant reduction in the critical value as the dimension is decreased to $5$.
We also study the theory with $N=1$, which has a $Z_2$ symmetry under $\phi\rightarrow -\phi$. We show that it possesses an
IR stable fixed point at imaginary couplings which can be reached by flow from a nearby fixed point describing a pair of $N=0$ theories.
We calculate certain operator scaling dimensions at the IR fixed point of the $N=1$ theory and suggest that, upon continuation to two
dimensions, it describes a non-unitary conformal minimal model.
}

\date{}
\maketitle

\tableofcontents

\section{Introduction and Summary}

This paper is a sequel to \cite{Fei:2014yja} where a one loop analysis was carried out for the cubic $O(N)$ symmetric theory of $N+1$ scalar fields
$\sigma$ and $\phi^i$ in $6-\epsilon$ dimensions. The Lagrangian of this theory is
\be
{\cal L}=\frac{1}{2}(\Par \phi^i)^2 + \frac{1}{2}(\Par \sigma)^2+ \frac{1}{2}g_1 \sigma (\phi^i)^2 + \frac{1}{6}g_2 \sigma^3
\ ,
\label{onmodels}
\ee
and the one loop beta functions showed that for $N>N_{\rm crit}$ there exists an IR stable fixed point with real values of the two couplings.
It was argued that this IR fixed point of the cubic $O(N)$ theory is equivalent to the perturbatively unitary UV fixed point of the $O(N)$ model with interaction
$(\phi^i \phi^i)^2$, which exists for large $N$ in
$4<d<6$ \cite{Parisi:1975im,parisi1977non,Bekaert:2011cu,Bekaert:2012ux}.
The $1/N$ expansions of various operator scaling dimensions were found in \cite{Fei:2014yja} to agree with the corresponding results
\cite{Vasiliev:1981yc,Vasiliev:1981dg,Vasiliev:1982dc,Lang:1990ni, Lang:1991kp, Lang:1992pp, Lang:1992zw, Petkou:1994ad,Petkou:1995vu}
in the quartic $O(N)$ model continued to $6-\epsilon$ dimensions.

A surprising result of \cite{Fei:2014yja} was that the one-loop value of $N_{\rm crit}$ is very large: if $N_{\rm crit}$ is treated as a continuous real parameter, then it is
$\approx 1038.266$.
Our main interest is in continuing the $d=6-\epsilon$ fixed point to $\epsilon=1$ in the hope of finding a 5-dimensional $O(N)$ symmetric unitary CFT.
In order to study the $\eps$ expansion of $N_{\rm crit}$, in section \ref{threeloops} we calculate the three-loop $\beta$ functions, following the earlier work of
\cite{deAlcantaraBonfim:1980pe,deAlcantaraBonfim:1981sy,Barbosa:1986kv}.\footnote{These papers considered cubic field theories of
$q-1$ scalar fields that were shown in \cite{Zia:1975ha} to describe the $q$-state Potts model. These theories possess only discrete symmetries and
generally differ from the $O(N)$ symmetric theories that we study.}
In section \ref{Ncrit-ep} we find the following expansion for the critical value of $N$:
\be
N_{\rm crit} = 1038.266 -609.840 \epsilon -364.173 \epsilon^2
+ {\cal O} (\epsilon^3).
\label{mainresult}\ee
Neglecting further corrections, this gives $N_{\rm crit}(\epsilon=1)\approx 64$, but higher orders in $\epsilon$ can obviously change this value
significantly. It is our hope that a conformal bootstrap approach \cite{Polyakov:1974gs,Ferrara:1973yt,Rattazzi:2008pe,Rychkov:2011et}, perhaps along the lines of \cite{Nakayama:2014yia}, can help determine $N_{\rm crit}$ more precisely in $d=5$. The bootstrap approach may also be applied in non-integer
dimensions close to $6$, but one should keep in mind that such theories are not strictly unitary \cite{Hogervorst:2014rta}.\footnote{
Another possible non-perturbative approach to the theory in $4<d<6$ is the Exact Renormalization Group \cite{Percacci:2014tfa}. This approach does not seem to indicate
the presence of a UV fixed point in the theory of $N$ scalar fields, but a search for an IR fixed point in the theory of $N+1$ scalar fields has not been carried
out yet.}

The major reduction of $N_{\rm crit}$ as $\epsilon$ is increased from $0$ to $1$ is analogous to what is known about the
Abelian Higgs model in $4-\eps$ dimensions.\footnote{
We are grateful to Igor Herbut for pointing this out to us.}
For the model containing $N_f$ complex scalars, the one-loop critical value of $N_f$ is found to be large,
$N_{f, \rm crit}\approx 183$ \cite{Halperin:1973jh}. However, the ${\cal O}(\epsilon)$ correction found from two-loop beta functions
has a negative coefficient and almost exactly cancels the leading term when $\epsilon=1$, suggesting that the $N_{f, \rm crit}$ is small
in the physically interesting three-dimensional theory \cite{herbut1997herbut}.

Another interesting property of the theories (\ref{onmodels}) is the existence of the lower critical value $N'_{crit}$ such that
for $N< N'_{crit}$ there is an IR stable fixed point at {\it imaginary} values of $g_1$ and $g_2$. The simplest example of
such a non-unitary theory is $N=0$, containing only the field $\sigma$. Its $6-\eps$ expansion was originally studied by Michael Fisher
\cite{Fisher:1978pf} and the continuation to $\epsilon=4$ provides
an approach to the Yang-Lee
edge singularity in the two-dimensional Ising model (this is the $(2,5)$ minimal model \cite{Belavin:1984vu,Cardy:1985yy} with central charge $-22/5$). From the three-loop $\beta$ functions we find the $\eps$ expansion
\be
N'_{\rm crit} = 1.02145 + 0.03253 \eps - 0.00163 \eps^2
+ {\cal O} (\epsilon^3)\ .
\ee
The smallness of the coefficients suggests that $N'_{crit}>1$ for a range of dimensions below $6$. In section \ref{nonunitary}
we discuss some properties of the $N=1$ theory. We show that it possesses an unstable fixed point with $g_1^*=g_2^*$ where the lagrangian splits into that
of two decoupled $N=0$ theories. There is also an IR stable fixed point where $g_2^*= 6 g_1^*/5 + O(\eps)$.
A distinguishing feature of this non-unitary CFT is that it has a discrete $Z_2$ symmetry, and it would be interesting to search for
it using the conformal bootstrap methods developed in \cite{Gliozzi:2014jsa}.
We suggest that, when continued to two dimensions,
it describes the $(3,8)$ non-unitary conformal minimal model.

In Section \ref{unitaryall} we also discuss unstable unitary fixed points that are present in $6-\epsilon$ dimensions for all $N$.
For $N=1$ the fixed point has $g_1^*=-g_2^*$; it is $Z_3$ symmetric and describes the critical point of the 3-state Potts model
in $6-\eps$ dimensions \cite{Amit:1979ev}.\footnote{We are grateful to Yu Nakayama for valuable discussions on this issue.}

\section{Three-loop $\beta$-functions in $d=6-\epsilon$}
\label{threeloops}

The action of the cubic theory is
\begin{align}
S = \int d^{d}x \big(\frac{1}{2}(\partial_{\mu}\phi^{i}_{0})^{2}+\frac{1}{2}(\partial_{\mu}\sigma_{0})^{2}+\frac{1}{2}g_{1,0} \sigma_{0}\phi^{i}_{0}\phi^{i}_{0} + \frac{1}{6}g_{2,0} \sigma_{0}^{3} \big),
\end{align}
where $\phi_{0}^{i}$ and $\sigma_{0}$ are bare fields and $g_{1,0}$ and $g_{2,0}$ are bare coupling constants.\footnote{We do not include mass terms as
we are ultimately interested in the conformal theory. In the dimensional regularization that we will be using, mass terms are not generated if we set them to zero from the start.}
As usual, we introduce renormalized fields and coupling constants by
\begin{equation}
\begin{aligned}
&\sigma_{0}=Z_{\sigma}^{1/2}\sigma, \quad \phi^{i}_{0}= Z^{1/2}_{\phi}\phi^{i},\\
&g_{1,0} =  \mu^{\frac{\epsilon}{2}} Z_{g_1} Z_{\sigma}^{-1/2} Z_{\phi}^{-1}\,g_{1}\,,
\quad g_{2,0}= \mu^{\frac{\epsilon}{2}} Z_{g_2} Z_{\sigma}^{-3/2}\,g_2 \,.
\end{aligned}
\end{equation}
Here $g_1$, $g_2$ are the dimensionless renormalized couplings, and $\mu$ is the renormalization scale. We may write
\begin{equation}
Z_{\sigma} = 1+\delta_{\sigma}\,,\quad Z_{\phi} = 1+\delta_{\phi} \,,\quad Z_{g_1} = 1+\delta g_1 \,,\quad  Z_{g_2} = 1+\delta g_2
\end{equation}
so that, in terms of renormalized quantities, the action reads
\begin{eqnarray}
&&S = \int d^{d}x \big(\frac{1}{2}(\partial_{\mu}\phi^{i})^{2}+\frac{1}{2}(\partial_{\mu}\sigma)^{2}
+\frac{g_{1}}{2}\sigma\phi^{i}\phi^{i} + \frac{g_{2} }{6}\sigma^{3} \\
&&~~~~~~~~~~~~~~~
+ \frac{\delta_{\phi}}{2}(\partial_{\mu}\phi^{i})^{2} +
\frac{\delta_{\sigma}}{2}(\partial_{\mu}\sigma)^{2} +\frac{\delta g_{1}}{2} \sigma \phi^{i}\phi^{i}  +\frac{\delta g_{2}}{6}\sigma^{3}\big). \notag
\end{eqnarray}
To carry out the renormalization procedure, we will use dimensional regularization \cite{'tHooft:1972fi} in $d=6-\epsilon$ and employ the minimal subtraction scheme \cite{'tHooft:1973mm}.
In this scheme, the counterterms are fixed by requiring cancellation of poles in the dimensional regulator, and have the structure
\begin{align}
\delta g_{1} =\sum_{n=1}^{\infty} \frac{a_{n }(g_{1},g_{2})}{\epsilon^{n}}, \quad \delta g_{2} = \sum_{n=1}^{\infty}\frac{b_{n }(g_{1},g_{2})}{\epsilon^{n}} , \quad
\delta_{\phi} =\sum_{n=1}^{\infty} \frac{z^{\phi}_{n}(g_{1},g_{2})}{\epsilon^{n}}, \quad \delta_{\sigma} =\sum_{n=1}^{\infty} \frac{z^{\sigma}_{n}(g_{1},g_{2})}{\epsilon^{n}}.
\end{align}
The anomalous dimensions and $\beta$-functions are determined by the coefficients of the simple $1/\epsilon$ poles in the counterterms
\cite{'tHooft:1973mm}. Specifically, in our case we have that the anomalous dimensions are given by
\begin{align}
\gamma_{\phi} &= -\frac{1}{4}\big(g_1 \frac{\partial}{\partial g_1} + g_2 \frac{\partial}{\partial g_2} \big) z_1^{\phi}, \\
\gamma_{\sigma} &= -\frac{1}{4}\big(g_1 \frac{\partial}{\partial g_1} + g_2 \frac{\partial}{\partial g_2} \big) z_1^{\sigma}
\end{align}
and the $\beta$-functions are
\begin{align}
&\beta_{1}(g_{1},g_{2}) = -\frac{\epsilon}{2}g_{1} + \frac{1}{2}\big(g_{1} \frac{\partial}{\partial g_{1}} +g_{2}\frac{\partial}{\partial g_{2}}-1\big)(a_{1}-\frac{1}{2}g_{1}(2 z_{1}^{\phi}+z_{1}^{\sigma})), \notag\\
&\beta_{2}(g_{1},g_{2}) = -\frac{\epsilon}{2}g_{2} + \frac{1}{2}\big(g_{1} \frac{\partial}{\partial g_{1}} +g_{2}\frac{\partial}{\partial g_{2}}-1\big)(b_{1}-\frac{3}{2}g_{2} z_{1}^{\sigma}).
\end{align}
In other words, in order to determine the anomalous dimensions and $\beta$-functions, we have to calculate the coefficients
of the $1/\epsilon$-divergencies in the loop diagrams, from which we can read off the residues
$a_1(g_1,g_2), b_1(g_2,g_2),z_{1}^{\phi}(g_1,g_2),z_{1}^{\sigma}(g_1,g_2)$.

Working in perturbation theory, we will denote by $a_{1i}$ the term
of $i$-th order in the coupling constants, and similarly for the other residue functions. Then, using the results for the Feynman diagrams collected
in the Appendix, we find the anomalous dimensions
\begin{equation}
\begin{aligned}
\gamma_{\phi} =& -\frac{1}{2}z_{12}^{\phi} - z_{14}^{\phi}-\frac{3}{2}z_{16}^{\phi} \\
=&\frac{g_1^2}{6(4\pi)^3} -  \frac{g_1^2}{432(4\pi)^{6}}  \left(g_1^2 (11 N-26)-48 g_1 g_2 +11 g_2^2\right)\\
&-\frac{g_{1}^{2}}{31104(4\pi)^9}\big( g_1^4(N(13N-232)+5184\zeta(3)-9064) +g_1^3 g_2 6(441N-544) \\
&-2g_1^2 g_2^2(193N-2592\zeta(3)+5881)+ 942g_1 g_2^3+327g_2^4\big),
\end{aligned}
\end{equation}
\begin{equation}
\begin{aligned}
\gamma_{\sigma} =&- \frac{1}{2}z_{12}^{\sigma} - z_{14}^{\sigma}-\frac{3}{2}z_{16}^{\sigma} \\
=&\frac{Ng_1^2+g_2^2}{12(4\pi)^3} +\frac{1}{432(4\pi)^{6}} \left(2N g_1^4 +48 N  g_1^3 g_2-11 N g_1^2 g_2^2+13 g_2^4\right) \\
&+\frac{1}{62208(4\pi)^9}\big( 96N(12N+11)g_1^5 g_2 - 1560 N g_1^3g_2^3+952N g_1^2 g_2^4-2Ng_1^6(1381N-2592\zeta(3)+4280)\\
&+g_2^6(2592\zeta(3)-5195)+3Ng_1^4 g_2^2(N+4320 \zeta(3)-8882)\big)
\end{aligned}
\end{equation}
and the $\beta$-functions
\begin{equation}
\begin{aligned}
\beta_{1}=&
-\frac{\epsilon}{2}g_{1} + (a_{13}-\frac{1}{2}g_{1}(2z_{12}^{\phi}+z_{12}^{\sigma}))
+2(a_{15}-\frac{1}{2}g_{1}(2z_{14}^{\phi}+z_{14}^{\sigma}))
+3(a_{17}-\frac{1}{2}g_{1}(2z_{16}^{\phi}+z_{16}^{\sigma}))\\
=&-\frac{\epsilon}{2}g_{1} +\frac{1}{12(4\pi)^{3}} g_{1}  \left((N-8)g_1^2-12g_1 g_2+g_2^2 \right)\\
&-\frac{1}{432(4\pi)^{6}} g_{1}  \left((536+86N)g_1^4+12(30-11N)g_1^3 g_2+(628+11N)g_1^2 g_2^2+24 g_1 g_2^3-13 g_2^4 \right) \\
&+\frac{1}{62208(4\pi)^9}g_1 \bigg(g_2^6(5195-2592\zeta(3))+12g_1 g_2^5(-2801+2592 \zeta(3)) \\
&-8g_1^2 g_2^4(1245+119N+7776 \zeta(3))+g_1^4 g_2^2(-358480+53990N-3N^2-2592(-16+5N)\zeta(3))\\
&+36g_1^5 g_2 (-500-3464N + N^2 + 864(5N-6)\zeta(3))\\
&-2g_1^6(125680-20344N+1831N^2+2592(25N+4)\zeta(3))+48g_1^3 g_2^3(95N-3(679+864\zeta(3)))\bigg),
\label{beta1}
\end{aligned}
\end{equation}
\begin{equation}
\begin{aligned}
\beta_{2}=&
-\frac{\epsilon}{2}g_{2} + (b_{13}-\frac{3}{2}g_{2} z_{12}^{\sigma})+2(b_{15}-\frac{3}{2}g_{2} z_{14}^{\sigma})
+3(b_{17}-\frac{3}{2}g_{2} z_{16}^{\sigma})  \\
=&-\frac{\epsilon}{2}g_{2} +\frac{1}{4(4\pi)^{3}} \left(-4Ng_1^3+Ng_1^2 g_2 -3g_2^3 \right)\\
&+\frac{1}{144(4\pi)^{6}} \left( -24Ng_1^5-322N g_1^4 g_2 - 60N g_1^3g_2^2+31N g_1^2 g_2^3-125g_2^5 \right)\\
&+\frac{1}{20736(4\pi)^9}\bigg(-48N(713+577N)g_1^7+6272N g_1^2 g_2^5+48N g_1^3 g_2^4(181+432\zeta(3))\\
&-5g_2^7(6617+2592\zeta(3))-24Ng_1^5 g_2^2(1054+471N+2592\zeta(3))\\
&+2Ng_1^6 g_2(19237N - 8(3713+324\zeta(3)))+3Ng_1^4g_2^3(263N-6(7105+2448\zeta(3))\bigg). \label{beta2}
\end{aligned}
\end{equation}
In the case $N=0$ (the single scalar cubic theory), our results are in agreement with the three-loop calculation of \cite{deAlcantaraBonfim:1981sy}.
\section{The IR fixed point}
Let us introduce the notation
\begin{align}
g_1 &\equiv \sqrt{\frac{6\eps(4\pi)^3}{N}}x, \quad  g_2 \equiv \sqrt{\frac{6\eps(4\pi)^3}{N}}y.
\label{gtoxy}
\end{align}
In terms of the new variables $x$ and $y$, the condition that both $\beta$-functions be zero reads
\begin{align}
 0=& \frac{1}{2}x(-8x^2+N(x^2-1)-12xy+y^2)\notag \\
&-\frac{1}{12N}x\left((536+86N)x^4+12(30-11N)x^3y+(628+11N)x^2y^2+24xy^3-13y^4\right) \eps \notag \\
&-\frac{1}{288N^2} x \big(12xy^5(2801-2592 \zeta(3))+y^6(-5195+2592\zeta(3))+48x^3y^3(2037-95N+2592\zeta(3))\notag\\
&+8x^2y^4(1245+119N+7776\zeta(3))+x^4y^2(358480-53990N+3N^2+2592(5N-16)\zeta(3))\notag\\
&-36x^5y(-500-3464N+N^2+864(5N-6)\zeta(3))\notag\\
&+2x^6(125680-20344N+1831N^2+2592(25N+4)\zeta(3))\big)\eps^2
\label{b1xy}
\end{align}
and
\begin{align}
0=&-\frac{1}{2}\left(9y^3+N(12x^3+y-3x^2y) \right) \notag\\
&-\frac{1}{4N}\left(125y^5+Nx^2(24x^3+322x^2y+60xy^2-31y^3) \right)\eps\notag\\
&-\frac{1}{96N^2} \big( N^2 x^4(27696x^3-38474x^2y+11304xy^2-789y^3)+5y^7(6617+2592\zeta(3))\notag\\
&+34224Nx^7-6272Nx^2y^5+16Nx^6y(3713+324\zeta(3))-48Nx^3y^4(181+432\zeta(3))\notag\\
&+48Nx^5y^2(527+1296\zeta(3))+18Nx^4y^3(7105+2448\zeta(3))\big) \eps^2.
\label{b2xy}
\end{align}
These equations can be solved order by order in the $\epsilon$ expansion. Using also the $1/N$ expansion, we find the fixed point values
\begin{align}
x_* &= 1 + \frac{22}{N}+\frac{726}{N^2}-\frac{326180}{N^3} - \frac{349658330}{N^4} +... \notag\\
&\qquad + \left( - \frac{155}{6N} - \frac{1705}{N^2} + \frac{912545}{N^3} + \frac{3590574890}{3N^4}+...\right)\eps \notag\\
&\qquad + \left(\frac{1777}{144N}+\frac{29093/36-1170\zeta(3)}{N^2}+...\right)\eps^2,
\label{xfixedpointlargeN}
\end{align}
\begin{align}
y_* &= 6(1 + \frac{162}{N}+\frac{68766}{N^2}+\frac{41224420}{N^3} + \frac{28762554870}{N^4} + ... \notag\\
&\qquad + \left( -\frac{215}{2N}-\frac{86335}{N^2}-\frac{75722265}{N^3}-\frac{69633402510}{N^4}+...\right)\eps \notag\\
&\qquad + \left(\frac{2781}{48N}+\frac{270911-157140\zeta(3)}{6N^2}+... \right)\eps^2.
\label{yfixedpointlargeN}
\end{align}
This large $N$ solution corresponds to an IR stable fixed point and generalizes the one-loop result of \cite{Fei:2014yja}. This fixed point
exists and is stable to all orders in the $1/N$ expansion.

If results beyond the $1/N$ expansion are desired, one can determine the the $\epsilon$ expansions of $x_*, y_*$ for finite $N$ as follows. Plugging the expansions
\begin{equation}
x_* = x_0(N)+x_1(N)\epsilon+x_2(N)\epsilon^2+\ldots\,,\qquad
y_* = y_0(N)+y_1(N)\epsilon+y_2(N)\epsilon^2+\ldots
\label{xy-eps}
\end{equation}
into (\ref{b1xy})-(\ref{b2xy}), the leading order terms are found to be \cite{Fei:2014yja,Giombi:2014xxa}
\begin{equation}
x_0(N) = \sqrt{\frac{N}{(N-44)z(N)^2+1}}z(N)\,,\qquad y_0(N) = \sqrt{\frac{N}{(N-44)z(N)^2+1}}(1+6z(N))\,,
\end{equation}
where $z(N)$ is the solution to the cubic equation
\begin{equation}
840z^3-(N-464)z^2+84z+5 = 0
\end{equation}
with large $N$ behavior $z(N) = 840N+O(N^0)$\footnote{The other two roots have large $N$ behavior $z(N)\sim \pm \sqrt{5N}$ and they are unstable IR fixed points \cite{Fei:2014yja}.}. This solution is real only if $N > 1038.27$, as can be seen from the discriminant of the above cubic equation. Once the $x_0(N), y_0(N)$ are known, one can then determine the higher order terms in (\ref{xy-eps}) by solving the equations (\ref{b1xy})-(\ref{b2xy}) order by order in $\epsilon$.

For $N\gg 1038$ the finite $N$ exact results are close to (\ref{xfixedpointlargeN})-(\ref{yfixedpointlargeN}), but for $N\sim 1038$ they deviate somewhat, indicating that, close to the critical $N$, the large $N$ expansion is not a good approximation (see also Figure \ref{GammaSigmavsN} below).

\subsection{Dimensions of $\phi$ and $\sigma$}
In terms of the rescaled couplings $x$, $y$ defined in (\ref{gtoxy}), the anomalous dimensions read
\begin{align}
\gamma_{\phi} =& \frac{x^2}{N}\eps - \frac{x^2}{12 N^2}\left((26-11N)x^2+48xy-11y^2 \right)\eps^2 \notag \\
&+\frac{x^2}{144N^3}\big(6(544-441N)x^3y-942xy^3-327y^4+x^4(9064+(232-13N)N-5184\zeta(3))\notag\\
&+2x^2y^2(5881+193N-2592\zeta(3))\}\big) \eps^3,\\
\gamma_{\sigma} =& \frac{Nx^2+y^2}{2N}\eps - \frac{1}{12N^2}\left(13y^4+Nx^2(2x^2+48xy-11y^2) \right)\eps^2 \notag\\
&+\frac{1}{288N^3}\bigg( N^2x^4(2762x^2-1152xy-3y^2)\notag\\
&+2N x^2\left(-528 x^3y+780xy^3-476y^4+3x^2y^2(4441-2160\zeta(3))+8x^4(535-324\zeta(3))\right)\notag\\
&+y^6(5195-2592\zeta(3))\bigg)\eps^3.
\label{anom-dim}
\end{align}
Plugging the fixed point values (\ref{xfixedpointlargeN})-(\ref{yfixedpointlargeN}) into these expressions, we get the conformal dimensions of $\sigma$ and $\phi$ at the fixed point
\begin{align}
\Delta_{\phi}&=\frac{d}{2}-1+\gamma_{\phi} \\
&=2-\frac{\epsilon}{2}+\left( \frac{1}{N} + \frac{44}{N^2} + \frac{1936}{N^3}+... \right) \eps
+ \left(-\frac{11}{12N} - \frac{835}{6N^2} - \frac{16352}{N^3} + ...\right) \eps^2\notag\\
&\quad + \left(-\frac{13}{144N}+\frac{6865}{72N^2}+\frac{54367/2-3672\zeta(3)}{N^3}+...\right)\eps^3, \\
\Delta_{\sigma}&=\frac{d}{2}-1+\gamma_{\sigma} \\
&=  2+\left(\frac{40}{N} + \frac{6800}{N^2} +... \right) \eps
+ \left(-\frac{104}{3N} - \frac{34190}{3N^2}  + ...\right) \eps^2 \notag\\
&\quad+\left(-\frac{22}{9N}+\frac{47695/18-2808\zeta(3)}{N^2}+...\right)\eps^3. \label{delphi-delsig}
\end{align}
One can verify that these results are in precise agreement with the large $N$ calculation of \cite{Vasiliev:1981yc,Vasiliev:1981dg,Vasiliev:1982dc} for the critical $O(N)$ model in general $d$, analytically continued to $d=6-\eps$. This provides a strong check on our calculations and on our interpretation of the IR fixed point
of the cubic $O(N)$ scalar theory.

\begin{figure}[t]
\centering
\includegraphics[width=12cm]{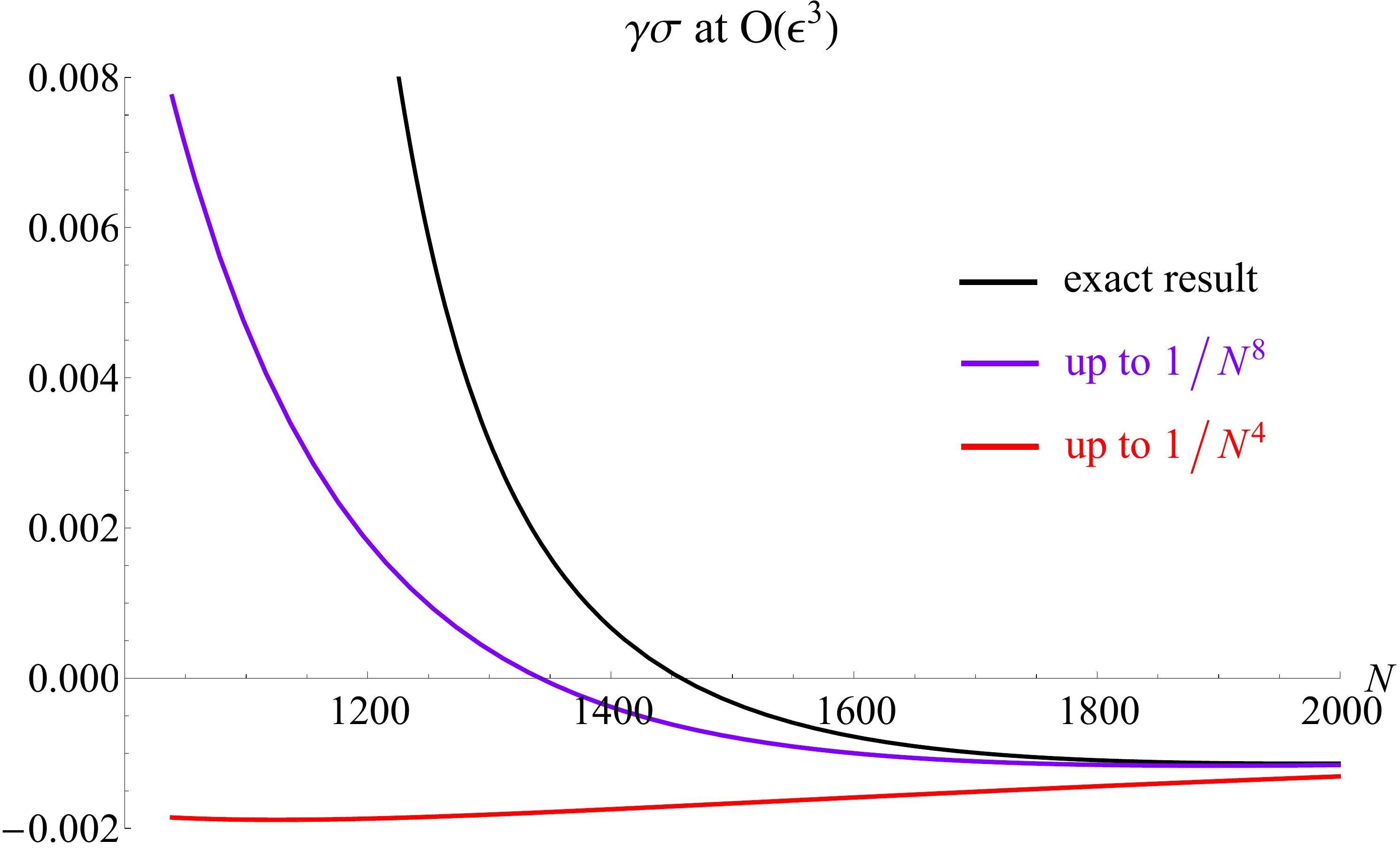}
\caption{The $O(\eps^3)$ in $\Delta_{\sigma}$ as a function of $N$ for $N \geq 1039$. The $1/N$ expansion approaches the exact result as we include more terms.}
\label{GammaSigmavsN}
\end{figure}

The $1/N$ expansions are expected to work well for $N\gg 1038$. For any $N$ larger than the critical value, the $\epsilon$ expansions of the scaling dimensions
may be determined using (\ref{anom-dim}) and the exact analytic solutions for the fixed point location $(x_*, y_*)$.
For example, in Figure \ref{GammaSigmavsN} we plot the coefficient of the $O(\eps^3)$ term in $\Delta_\sigma$ as a function of $N$ and compare it with the corresponding $1/N$ expansion.

\subsection{Dimensions of quadratic and cubic operators}

In \cite{Fei:2014yja} the mixed anomalous dimensions of quadratic operators $\sigma^2$ and $\phi^i\phi^i$ were calculated at one-loop order. These results were checked against the $O(1/N)$ term in the corresponding operator dimensions for the $O(N)$ $\phi^4$ theory \cite{Lang:1992zw}. In this paper, we carry out an additional check,
comparing with the $O(1/N^2)$ correction found in \cite{Broadhurst:1996ur}, but still working to the one-loop order in $\epsilon$ (it should be straightforward
to generalize the mixing calculation to higher loops, but we will not do it here).

In the quartic $O(N)$ theory with interaction $\frac \lambda 4 (\phi^i\phi^i)^2$, the derivative of the beta function at the fixed point coupling
\begin{equation}
\omega = \beta'(\lambda_*)=4-d+\frac{\omega_1}{N}+\frac{\omega_2}{N^2}+\ldots
\end{equation}
is related to the dimension of the operator $(\phi^i\phi^i)^2$ by
\begin{equation}
\Delta_{\phi^4} = d+\omega\,.
\end{equation}
In \cite{Lang:1992zw,Broadhurst:1996ur} the coefficient $\omega_1$ was computed as a function of dimension $d$:
\begin{equation}
\omega_1 = \frac{2 (d-4) (d-2) (d-1) \Gamma\left(d\right)}{d \Gamma \left(2-\frac{d}{2}\right) \Gamma \left(\frac{d}{2}\right)^3}\ .
\label{om1}
\end{equation}
The coefficient $\omega_2$ has a more complicated structure for general $d$ which was first found in \cite{Broadhurst:1996ur}.
Using this result, we get that in $d=5$,
\begin{equation}
\Delta_{\phi^4} = 4- {2048\over 15 \pi^2 N}- {8192 ( 67125 \pi^2-589472)\over 3375 \pi^4 N^2}+\ldots
\approx 4 -{13.8337\over N}-{1819.66\over N^2} +\ldots
\end{equation}
Let us also quote the expansion of $\omega_2$ in $d=4-\epsilon$ and $d=6-\epsilon$:
\begin{align}
&\omega_2 = 102\epsilon^2+\left(-\frac{259}{2}+120\zeta(3)\right)\epsilon^3+\ldots \,,\qquad d=4-\epsilon \\
&\omega_2 = -49760 \epsilon+\frac{237476}{3}\epsilon ^2+\left(-\frac{92480}{9}+32616 \zeta (3)\right) \epsilon ^3+\ldots \,,\qquad d=6-\epsilon
\label{om2}
\end{align}
In $d=4-\epsilon$, one can check that the above results correctly reproduce the derivative of the $\beta$-function \cite{Wilson:1973jj}
\begin{equation}
\beta = -\epsilon \lambda +\frac{N+8}{8 \pi ^2}\lambda^2 -\frac{3 (3N+14)}{64 \pi ^4}\lambda^3+
\frac{33N^2+480 N \zeta(3)+ 922 N+ 2112 \zeta(3) +2960}{4096 \pi^6}\lambda^4+
O(\lambda^5)
\end{equation}
at the IR fixed point
\begin{equation}
\begin{aligned}
&\lambda_* = \frac{8 \pi ^2 }{N+8}\epsilon+\frac{24 \pi ^2 (3N+14)}{(N+8)^3}\epsilon^2\\
&~~~~~~
-\frac{\pi ^2 (33N^3 - 110 N^2 +96 (N+8) (5N+22) \zeta(3) -1760 N -4544)}{(N+8)^5}\epsilon^3
+O(\epsilon^4)
\ .
\end{aligned}
 \end{equation}

In $d=6-\epsilon$, the dimension of the  $(\phi^i\phi^i)^2$ operator in the quartic theory should be matched to the primary operator arising from the mixing of the $\sigma^2$ and $\phi^i\phi^i$ operators in our cubic theory.
In \cite{Fei:2014yja}, the mixing matrix of $\sigma^2$ and $\phi^i\phi^i$ to one-loop order was found to be
\begin{equation}
\gamma^{ij} = \frac{-1}{6(4\pi)^3} \left( \begin{array}{cc}
4 g_1^2 -N g_1^2& 6 \sqrt{N} g_1^2 - \sqrt{N}g_1g_2 \\
6 \sqrt{N} g_1^2 - \sqrt{N}g_1g_2 & 4g_2^2-Ng_1^2  \end{array} \right).
\end{equation}
Computing the eigenvalues $\gamma_{\pm}$ of this matrix, and inserting the values of one-loop fixed point couplings
\begin{align}
g_{1*}&=\sqrt{\frac{6\eps(4\pi)^3}{N}}\left( 1+\frac{22}{N}+\frac{726}{N^2}-\frac{326180}{N^3}+\ldots \right), \\
g_{2*}&=6 \sqrt{\frac{6\eps(4\pi)^3}{N}} \left(1+\frac{162}{N}+\frac{68766}{N^2}+\frac{41224420}{N^3}+\ldots \right)
\end{align}
we find the scaling dimensions of the quadratic operators to be
\begin{align}
\Delta_{-}&=d-2+\gamma_{-}=4+\left(-\frac{100}{N}-\frac{49760}{N^2}-\frac{27470080}{N^3} +\ldots \right)\eps + \mathcal{O}(\eps^2),  \\
\Delta_{+}&=d-2+\gamma_{+}=4+\left(\frac{40}{N}+\frac{6800}{N^2}+\frac{2637760}{N^3} +\ldots \right)\eps + \mathcal{O}(\eps^2).
\end{align}
The operator with dimension $\Delta_{+}$ is a descendant of $\sigma$. The operator with dimension $\Delta_{-}$ is a primary, and comparing with (\ref{om1}), (\ref{om2}), we see that its
dimension precisely agrees with the results of \cite{Broadhurst:1996ur} to order $1/N^2$. The higher order terms in $\eps$ can be determined from mixed anomalous dimension calculations beyond one loop, and we leave this to future work.

We now calculate the mixed anomalous dimensions of the nearly marginal operators $\OO^1 = \sigma\phi\phi$ and $\OO^2 = \sigma^3$.
Using the beta functions written in equations (\ref{beta1})-(\ref{beta2}), we can determine the anomalous dimensions of the nearly marginal operators
by computing the eigenvalues of the matrix
\begin{equation}
M_{ij} = \frac{\partial \beta_i}{\partial g_j}\,.
\label{Mmatrix}
\end{equation}
Strictly speaking, this matrix is not exactly equal to the anomalous dimension mixing matrix, because it is not symmetric. However, we could make it
symmetric by dividing and multiplying the off-diagonal elements by a factor $\sqrt{3N}+O(\epsilon)$, which corresponds to an appropriate rescaling of the
couplings. This clearly does not change the eigenvalues of the matrix, and hence we can directly compute the eigenvalues $\lambda_{\pm}$ of (\ref{Mmatrix}),
and obtain the dimensions of the eigenstate operators as
\begin{equation}
\Delta_{\pm}=d+\lambda_{\pm}\, .
\label{margindim}
\end{equation}
Plugging in the fixed point values $x_*$ and $y_*$ from equations (\ref{xfixedpointlargeN})-(\ref{yfixedpointlargeN}), we find that
\begin{equation}
\begin{aligned}
\label{nearlymarginal}
&\Delta_{+} = 6 + \left(\frac{155}{3}\eps^2 -\frac{1777}{36} \eps^3 + ... \right)\frac{1}{N} + \mathcal{O}(\frac{1}{N^2})\ ,\\
&\Delta_{-} = 6 + \left( -420\eps + 499 \eps^2 - \frac{1051}{12} \eps^3 +...\right)\frac{1}{N} + \mathcal{O}(\frac{1}{N^2}).
\end{aligned}
\end{equation}
The dimension of the $\sigma^k$ operator in the quartic $O(N)$ model is known to order $1/N$ as function of $d$ \cite{Lang:1992zw}, and
may be written as
\begin{equation}
\Delta(\sigma^k)= 2k
+ \frac{k (d-2) \big ((k-1)d^2-d(3k-1) +4 \big ) \Gamma\left(d\right)}{N d \Gamma \left(2-\frac{d}{2}\right) \Gamma \left(\frac{d}{2}\right)^3}
+ \mathcal{O}(\frac{1}{N^2})
\ .
\end{equation}
Our result for $\Delta_{-}$ agrees with the $\epsilon$ expansion of this formula for $k=3$ in $d=6-\epsilon$.

\section{Analysis of critical $N$ as a function of $\epsilon$}
\label{Ncrit-ep}

We now investigate the behavior of $N_{\rm crit}$ above which the fixed point exists at real values of the couplings. This can be defined as the value of $N$ (formally viewed as a continuous parameter) at which two real solutions of the $\beta$-function equations merge, and subsequently go off to the complex plane. Geometrically, this means that the curves on the $(g_{1},g_{2})$ plane defined by the zeroes of $\beta_1$ and $\beta_2$ are barely touching, i.e. they are tangent to each other. Therefore the critical $N$,
as well as the corresponding critical
value of the couplings, can be determined by solving the system of equations
\begin{equation}
\begin{aligned}
&\beta_1 = 0\,,\qquad \beta_2=0, \\
&\frac{\partial \beta_1/ \partial g_1}{\partial \beta_1/ \partial g_2} = \frac{\partial \beta_2/ \partial g_1}{\partial \beta_2/ \partial g_2}.
\label{Ncrit-system}
\end{aligned}
\end{equation}
Note that the condition in the second line is equivalent to requiring that the determinant of the anomalous dimension mixing matrix of nearly marginal operators, $M_{ij}=\frac{\partial \beta_i}{\partial g_j}$, vanishes. This means that one of the two eigenstates becomes marginal.

Working in terms of the rescaled coupling constants defined in (\ref{gtoxy}), we can solve the system of equations (\ref{Ncrit-system}) order by order in $\epsilon$.
We assume a perturbative expansion
\begin{equation}
\begin{aligned}
x&=x_0 + x_1 \eps + x_2 \eps^2 +O(\eps^3), \\
y&=y_0 + y_1 \eps + y_2 \eps^2 +O(\eps^3),\\
N&=N_0 +N_1 \eps + N_2 \eps^2+O(\eps^3)
\label{small-eps-crit}
\end{aligned}
\end{equation}
and plugging this into (\ref{Ncrit-system}), we can solve for the undetermined coefficients uniquely. At the zeroth order, we get the equations
\begin{equation}
\begin{aligned}
&N_0 + 8x_0^2-N_0 x_0^2 + 12 x_0 y_0 - y_0^2 =0,\\
&12 N_0 x_0^3 +N_0 y_0 -3N_0 x_0^2y_0+9y_0^3 = 0,\\
&6+\frac{(N_0-44)x_0}{6x_0 - y_0} = \frac{6N_0 x_0 (y_0-6x_0)}{3N_0 x_0^2-27y_0^2-N_0}.
\label{Ncrit-zeroth}
\end{aligned}
\end{equation}
The above system of equations can be solved analytically, as was done in \cite{Fei:2014yja}. We find that, up to the signs of $x_0$ and $y_0$, there are three inequivalent solutions
\begin{align}
x_0 &= 1.01804 \,,\qquad y_0=8.90305 \,,\qquad N_0=1038.26605, \label{largeNsol} \\
x'_0 &= 0.23185 i \,,\qquad y'_0=0.25582 i \,,\qquad N'_0=1.02145,  \label{smallNsol} \\
x''_0 &= 0.13175 \,,\qquad y''_0=-0.03277 \,,\qquad N''_0=-0.08750. \label{negNsol}
\end{align}
\begin{figure}[t]
\centering
\includegraphics[width=12cm]{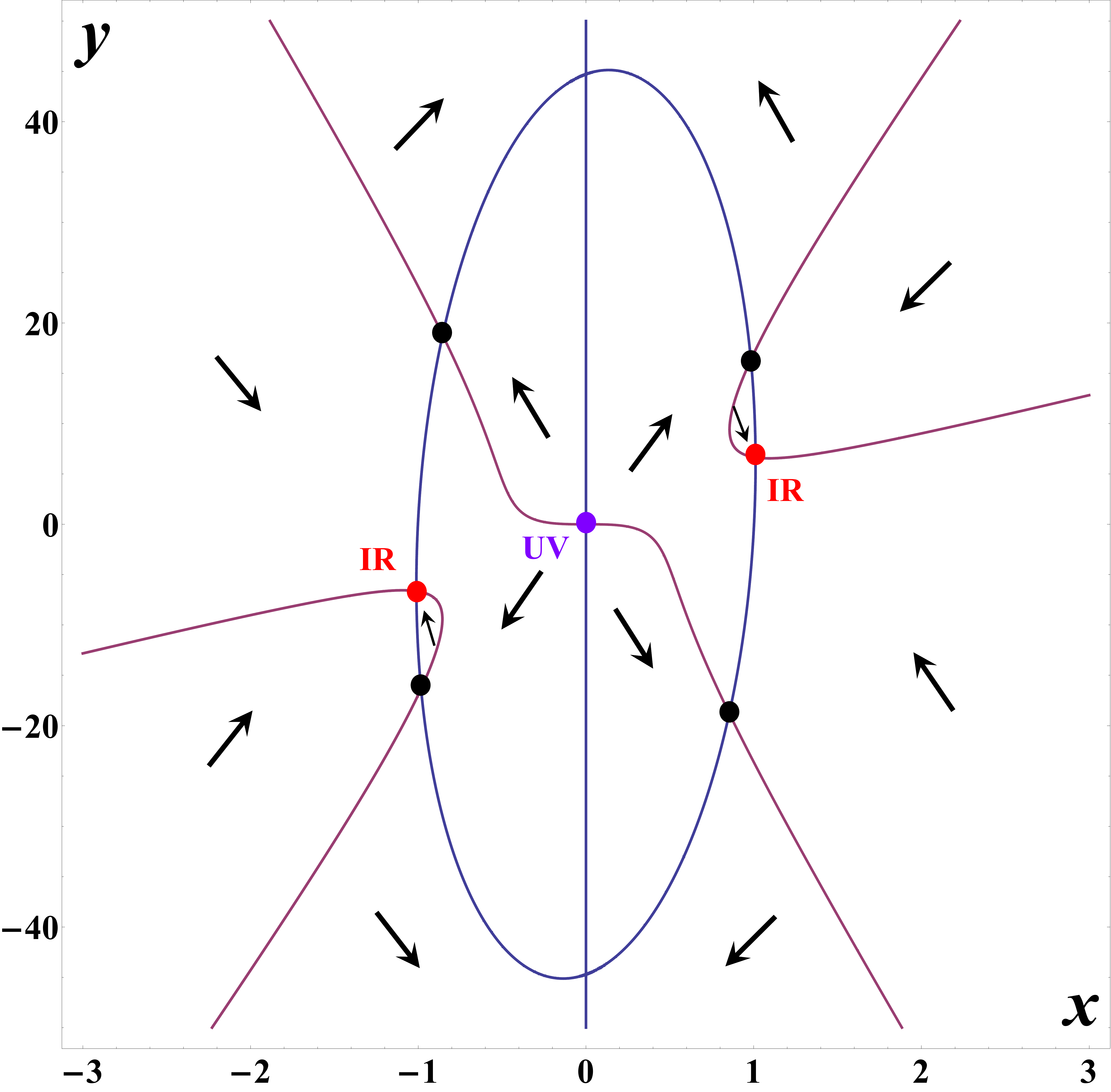}
\caption{The zeroes of the one loop $\beta$ functions and the RG flow directions for $N=2000$. The red dots correspond to the stable IR fixed points, while the black dots are unstable fixed points. As $N\rightarrow N_{\rm crit}$, the red dot merges with the nearby black dot, and the two fixed points move into the complex plane.}
\label{flow}
\end{figure}

The first of these solutions, with $N_{\rm crit}=1038.26605 + O(\epsilon)$, is of most interest to us because it is related to the large $N$ limit of the theory.
For $N>N_{\rm crit}$, we find a stable IR fixed point at real couplings $g_1$ and $g_2$.\footnote{It is stable with respect to flows of the nearly marginal
couplings $g_1$ and $g_2$. As usual, there are some $O(N)$ invariant relevant operators that render this fixed point not perfectly stable.}
This fixed point is shown with the red dot in Figure \ref{flow} (there is a second stable IR fixed point obtained by the transformation
$(g_1,g_2)\rightarrow (-g_1,-g_2)$, which is a symmetry of this theory). There is also a nearby unstable fixed point, shown with a black dot, which has one stable and one unstable direction. As $N$ approaches $N_{\rm crit}$ from above, the nearby unstable fixed point approaches the IR stable fixed point, and they merge at $N_{\rm crit}$. At $N < N_{\rm crit}$, both fixed points disappear into the complex plane. As discussed in \cite{Kaplan:2009kr}, this is a rather generic behavior at the lower edge of the conformal window: the conformality is lost through the annihilation of a UV fixed point and an IR fixed point. In \cite{Kaplan:2009kr} this was argued to happen at the lower (strongly coupled) edge of the conformal window for 4-dimensional $SU(N_c)$ gauge theory with $N_f$ flavors. It is interesting to observe that the same type of behavior occurs at
the lower edge of the conformal window of the $O(N)$ model in $d=6-\epsilon$, which extends from $N_{\rm crit}$ to infinity.

Let us identify the operator that causes the flow between the unstable fixed point and the IR stable fixed point of our primary interest.
It is one of the two nearly marginal operators cubic in the fields that were studied in section 3.2. By studying the behavior of the dimensions
$\Delta_1$ and $\Delta_2$ as $N\rightarrow N_{\rm crit}$ we find that $\Delta_2\rightarrow 6-\epsilon$. Therefore, it is the operator corresponding to
$\Delta_2$ that becomes exactly marginal for $N=N_{\rm crit}$ and causes the flow between the IR fixed point and the nearby UV fixed point for $N$
slightly above $N_{\rm crit}$. In bootstrap studies of the quartic $O(N)$ model this operator was denoted by $\sigma^3$ \cite{Lang:1992zw}, i.e. it can be thought of as
the ``triple-trace operator" $(\phi^i\phi^i)^3$. The theory at the unstable fixed point has an unconventional large $N$ behavior
where $x \sim \mathcal{O}(1)$ and $y \sim \mathcal{O}(\sqrt{N})$, so that corrections to scaling dimension proceed in powers of $N^{-1/2}$ \cite{Fei:2014yja}.

Let us now go back to finding the higher order corrections to $N_{\rm crit}$ given by (\ref{largeNsol}) (the higher order corrections
to the other critical values (\ref{smallNsol})-(\ref{negNsol}) will be discussed in the next section).
Once we have solved the leading order system (\ref{Ncrit-zeroth}), we can plug the solution into (\ref{small-eps-crit}) and expand (\ref{Ncrit-system}) up to order $\epsilon^2$.  From this we obtain simple systems of linear equations from which we can determine $x_1, y_1, N_1$ and $x_2, y_2, N_2$. We find
\begin{equation}
\begin{aligned}
&x_1 = -0.00940\,,\qquad
y_1 = -0.21024\,,\qquad
N_1 = -609.93980,\\
&x_2 = 0.00690\,,\qquad
y_2 = 1.01680\,,\qquad
N_2 = -364.17333.
\end{aligned}
\end{equation}
Thus, to three loop order, we conclude that
\begin{equation}
N_{{\rm crit}} = 1038.26605 - 609.83980 \eps - 364.17333 \eps^2+O(\eps^3)\,.
\end{equation}
We have also checked these expansion coefficients via a direct high-precision numerical calculation of $N_{\rm crit}$ for very small values of $\epsilon$.
The large and negative coefficients indicate that in the physically interesting case of $d=5$, $N_{\rm crit}$ is likely to be much lower than the
zeroth order value (this is analogous to the result \cite{herbut1997herbut} for the Abelian Higgs model). If we just use the first three terms and plug in $\eps=1$, we get:
\begin{equation}
N_{\rm crit} \approx 64.253\,.
\label{ncritfive}
\end{equation}

For $N<N_{\rm crit}$ the anomalous dimensions, such as $\gamma_\phi$, are no longer positive (in fact, they become complex).
This loss of positivity of $\gamma_\phi$ can also be seen as $N$ is reduced in the quartic $O(N)$ model.
For example, using the $1/N$ expansion of $\gamma_\phi$ in $d=5$ \cite{Vasiliev:1982dc}
\begin{eqnarray}
\gamma_{\phi} &=& \frac{32}{15\pi^2N}-\frac{1427456}{3375 \pi^4N^2} \cr
&+&\left(\frac{275255197696}{759375 \pi^6}-\frac{89735168}{2025 \pi^4}+\frac{32768 \ln 4}{9 \pi^4}-\frac{229376 \zeta(3)}{3\pi^6}\right)\frac{1}{N^3}+\ldots \cr
&=& \frac{3}{2}+\frac{0.216152}{N}-\frac{4.342}{N^2}-\frac{121.673}{N^3}+\ldots
\label{delphi-5d}
\end{eqnarray}
we find that it stops being positive for $N< 35$. This critical value is not too far from (\ref{ncritfive}).

It is also instructive to study the theory using the $4-\epsilon$ expansion.
The anomalous dimensions of $\phi^i$ is \cite{Wilson:1973jj}
\begin{eqnarray}
\gamma_\phi &=& {N+2\over 4 (N+8)^2} \epsilon^2 + {N+2\over 16 (N+8)^4} \left (-N^2 + 56 N +272\right )\epsilon^3 \cr
&+&  {N+2\over 64 (N+8)^6} \left (-5 N^4 -230 N^3+ 1124 N^2+17920 N + 46144 - 384 \zeta(3) ( 5 N +22) (N+8) \right )\epsilon^4\cr
& +& {\cal O}(\epsilon^5)
\end{eqnarray}
For positive $\epsilon$ this expansion gives accurate information about the Wilson-Fisher IR fixed points \cite{Wilson:1971dc}.
For negative $\epsilon$ there exist formal UV fixed points at negative quartic coupling where we can apply this formula as well.
In that case, $\gamma$ becomes negative for sufficiently large $|\epsilon|$ and $N<N_{\rm crit}$, indicating that the operator $\phi^i$
violates the unitarity bound. For example for $d=5$, corresponding to $\epsilon=-1$, we find $N_{\rm crit}\approx 8$. Inclusion of the
${\cal O}(\epsilon^5)$ term raises this to $N_{\rm crit}\approx 14$  (it is not clear, however, that this is a better estimate since the $\epsilon$ expansion is asymptotic
and the coefficient of $\epsilon^5$ is much larger than the previous ones).

We see, therefore, that the estimates of $N_{\rm crit}$ using the quartic $O(N)$ theory in $d=5$ are even lower than the three-loop estimate (\ref{ncritfive}).
It seems safe to conclude that the true value is much lower than the one-loop estimate of $1038$. To determine $N_{\rm crit}$ in $d=5$ more precisely,
one needs a non-perturbative
approach to the $d=5$ theory, perhaps along the lines of the conformal bootstrap calculation in \cite{Nakayama:2014yia}.

\subsection{Unitary fixed points for all positive $N$}
\label{unitaryall}

Let us note that not all real fixed points disappear for $N< N_{\rm crit}$. The unstable real fixed points that are located in the upper left and lower right corners of Figure \ref{flow} exist for all positive $N$, and we would like to find their interpretation.

The fixed point with $N=1$ has a particularly simple property that $g_1^*=-g_2^*$. This property of the solution holds for the three loop $\beta$ functions, and
we believe that it is exact. Using this, we note that the action at the fixed point is proportional to $(\sigma+i\phi)^3 + (\sigma-i\phi)^3$.
Therefore, the theory at this fixed point enjoys a $Z_3$ symmetry acting by the phase rotation on the complex combination $\sigma+i\phi$.
This cubic classical action appears in the Ginzburg-Landau theory
for the 3-state Potts model (see, for example, \cite{Amit:1979ev}).\footnote{We are grateful to Yu Nakayama for pointing this out to us.}
Therefore, we expect the $Z_3$ symmetric fixed point to
describe the 3-state Potts model in $d=6-\eps$.
The dimensions of operators at this fixed point are related by the $Z_3$ symmetry. For example, we find
\begin{equation}
\Delta_\phi= \Delta_\sigma=
2 - \frac {1} {3}\eps + \frac{2} {3}\eps^2 + \frac{443} { 54}\eps^3 + {\cal O}(\epsilon^4)
 \ .
\end{equation}
This is in agreement with the result of \cite{deAlcantaraBonfim:1981sy}.
By calculating the eigenvalues $\lambda_\pm$ of the matrix $M_{ij}=\frac{\partial \beta_i}{\partial g_j}$, we also find the dimensions
(\ref{margindim}) of
the two cubic operators to order $\epsilon^3$:
\begin{align}
\Delta_{-}&= 6 - \frac{14}{3}\eps - \frac{158} { 9}\eps^2 -\left (\frac{17380}{ 81}+ 16 \zeta(3)\right )\eps^3
=6- 4.66667 \eps - 17.5556 \eps^2 - 233.801 \eps^3 \ ,\nonumber \\
\Delta_{+}&= 6- \frac{83} { 18}\eps^2 -\left (\frac{38183}{ 648}+ 4 \zeta(3)\right )\eps^3=  6- 4.61111 \eps^2 - 63.7326  \eps^3\ .
\label{z3eps}
\end{align}
The dimension $\Delta_+$ corresponds to the operator  $(\sigma+i\phi)^3 + (\sigma-i\phi)^3$
which preserves the $Z_3$ symmetry and is slightly irrelevant for small $\epsilon$.
The dimension $\Delta_-$ corresponds to the relevant operator $\sigma (\sigma^2 + \phi^2)$ which breaks the $Z_3$.
Thus, the $Z_3$ symmetry helps stabilize the fixed point at small $\epsilon$.

Unfortunately, the $6-\epsilon$ expansions (\ref{z3eps}) have growing coefficients, and it is not clear for what range of $\epsilon$ the
fixed point exists. Thus, one may not be able to interpolate smoothly from the
$Z_3$ symmetric fixed point in $d=6-\epsilon$ to $d=2$ where the 3-state Potts model is described by the unitary
$(5,6)$ minimal model \cite{Belavin:1984vu}.

For all $N\geq 2$ we find unstable fixed points with $O(N)$ symmetry. These fixed points always have a relevant cubic operator, corresponding
to a negative eigenvalue of the matrix $\frac{\partial \beta_i}{\partial g_j}$. Also, they exhibit an unconventional large $N$ behavior
involving half-integer powers of $N$, similarly to the unstable fixed points that appear for
$N>N_{\rm crit}$ and are shown by the black dots in the upper right and lower left corners of Figure \ref{flow}.
We leave a discussion of these fixed points for the future.

\section{Non-unitary theories}
\label{nonunitary}

In addition to the fixed points studied so far, which are perturbatively unitary and appear for $N>N_{\rm crit}$, there exist non-unitary fixed points for
$N''_{\rm crit} < N < N'_{\rm crit}$. The leading values of $N'_{\rm crit}$ and $N''_{\rm crit}$ are given in (\ref{smallNsol}) and (\ref{negNsol}), respectively.
Using the method developed above for finding the higher order in $\eps$ corrections to $N_{\rm crit}$ we get
\begin{equation}
\begin{aligned}
&N'_{\rm crit} = 1.02145 + 0.03253 \eps - 0.00163 \eps^2\\
&x' = i\left(0.23185 + 0.08887 \eps -0.03956\eps^2\right)\,,\quad
y' = i\left(0.25582 + 0.11373 \eps -0.04276 \eps^2\right)
\label{lowercrit}
\end{aligned}
\end{equation}
and
\begin{equation}
\begin{aligned}
&N''_{\rm crit} = -0.08750 + 0.34726 \eps - 0.88274 \eps^2\\
&x'' = 0.13175 -0.16716 \eps +0.12072 \eps^2\,,\quad
y'' = -0.03277 + 0.13454 \eps -0.35980 \eps^2
\end{aligned}
\end{equation}
Unfortunately, the latter expansion has growing coefficients, and we cannot extract any useful information from it.
On the other hand, the higher order corrections to $N'_{\rm crit}$ are very small, which suggests that $N'_{\rm crit}>1$ for range of dimensions below $6$.

The theory with $N=0$, which contains only the field $\sigma$, was originally studied by Michael Fisher as an approach to the Yang-Lee
edge singularity in the Ising model \cite{Fisher:1978pf}.
Since the coupling is imaginary, it describes a non-unitary theory where some operator dimensions (e.g. $\sigma$) are below the unitarity bounds. In $d=2$, this CFT
corresponds to the $(2,5)$ minimal model \cite{Cardy:1985yy}, which has $c=-22/5$.
A conformal bootstrap approach to this model \cite{Gliozzi:2014jsa} has produced good results for a range of dimensions below $6$.

The $N=1$ theory, which has two fields and two coupling constants, has a more intricate structure. This theory is distinguished from the $N=0$ case by the
presence of a $Z_2$ symmetry $\phi \rightarrow -\phi$.
Examining the $\beta$ functions at $N=1$ and the eigenvalues of the matrix $\frac{\partial \beta_i}{\partial g_j}$, we observe that there exist a stable fixed point with
$g_2^* =6g_1^*/5+O(\epsilon)$, and an unstable one with $g_1^*=g_2^*$.
Introducing the field combinations
\begin{equation}
\sigma_1= \sigma+\phi\ , \quad \sigma_2=\sigma-\phi\ ,
\end{equation}
we note that
for $g_1^*=g_2^*$ the interactions of the $N=1$ model decouple as $\sim \sigma_1^3 + \sigma_2^3$,
i.e. at this fixed point the theory is a sum of two Fisher's $N=0$ theories.
However, one of the flow directions at this fixed point is unstable, since the corresponding operator has
$\Delta_{\cal O}= 6-10\eps/9 + O(\eps^2)$ and is relevant (this value of the dimension corresponds to the negative eigenvalue of the matrix
$M_{ij}=\frac{\partial \beta_i}{\partial g_j}$ at the $g_1^*=g_2^*$ fixed point). This dimension has a simple explanation as follows.
The flow away from the decoupled fixed point is generated by the operator ${\cal O}=\sigma_1 \sigma_2^2 + \sigma_2 \sigma_1^2$. This is
allowed by the original  $Z_2$ symmetry
$\phi\rightarrow -\phi$, which translates into the interchange of $\sigma_1$ and $\sigma_2$.
Thus,
\be
\label{oprelation}
\Delta_{\cal O}= \Delta_\sigma^{N=0}+ \Delta_{\sigma^2}^{N=0}= 2 + 2 \Delta_\sigma^{N=0}
\ ,
\ee
where we used the fact that in the $N=0$ theory, $\Delta_{\sigma^2}^{N=0}= 2+\Delta_\sigma^{N=0}$ because $\sigma^2$ is a descendant.
Using (\ref{anom-dim}) for $N=0$, we find
\be
\Delta_{\sigma} = 2 - {5\over 9}\eps -{43\over 1458}\eps^2+ \left ({8 \zeta(3)\over 243} -{8375\over 472392} \right ) \eps^3
=2- 0.555556 \eps - 0.0294925\eps^2 + 0.021845 \eps^3
\label{sigmadim}
\ee
Substituting this into (\ref{oprelation}) we find the dimension of the relevant operator ${\cal O}$, which indeed precisely agrees with $\Delta_{\cal O}=d+\lambda_{-}$, where
$\lambda_{-}$ is the negative eigenvalue of $M_{ij}=\frac{\partial \beta_i}{\partial g_j}$ at the $g_1^*=g_2^*$ fixed point.
Using the $\epsilon$ expansion (\ref{sigmadim}), we find that ${\cal O}$ continues to be relevant as $\epsilon$ is increased. For $\epsilon=4$, i.e. $d=2$,
we know the exact result in the $(2,5)$ minimal model that $\Delta_\sigma^{N=0}=-2/5$, which implies $\Delta_{\cal O}=6/5$. This strongly suggests that
${\cal O}$ is relevant, and the decoupled fixed point is unstable, for the entire range $2\leq d < 6$.
To describe this CFT in $d=2$ more precisely, we note the existence of the modular invariant minimal model $M(3,10)$, which is closely related to the product of two
Yang-Lee $(2,5)$ minimal models \cite{2011NJPh...13d5006A,Quella:2006de}.

The flow away from the unstable fixed point with $g_1^*=g_2^*$ can lead the $N=1$ theory to the IR stable fixed point where $g_2^*= 6 g_1^*/5 + O(\eps)$.
Using our results we can deduce the $\eps$ expansion of various operator dimensions at this fixed point.
For example,
\begin{align}
\Delta_{\phi}&=2- 0.5501 \eps -0.0234477 \eps^2+0.0200649 \eps^3 +\ldots \nonumber \\
\Delta_{\sigma}&=2-0.561122  \eps- 0.0358843  \eps^2 + 0.0236057  \eps^3 + \ldots
\label{dimnonunit}
\end{align}
By calculating the eigenvalues $\lambda_\pm$ of the matrix $M_{ij}=\frac{\partial \beta_i}{\partial g_j}$, we find the dimensions of
two operators that are slightly irrelevant in $d=6-\eps$
\begin{align}
\Delta_{-}&=d+\lambda_-= 6-0.88978 \eps + 0.0437732 \eps^2 - 0.039585 \eps^3\ ,\nonumber \\
\Delta_{+}&=d+\lambda_+=  6-0.773191 \eps^2 + 1.59707 \eps^3\ .
\end{align}
As $\epsilon$ is increased, these expansions suggest that the two operators become more irrelevant.
It would be interesting to study this $Z_2$ symmetric fixed point using a conformal bootstrap approach along the lines of \cite{Gliozzi:2014jsa}.

Assuming that the $N=1$ IR fixed point continues to be stable in $d=5,4, 3, 2$,
it is interesting to look for statistical mechanical interpretations of this non-unitary CFTs. A distinguishing feature of the $N=1$ CFT is that it has a discrete $Z_2$ symmetry, while the $N=0$ theory has no symmetries at all.
As we have noted, in $d=2$ the CFT can be obtained via deforming the $(3,10)$ minimal model by a Virasoro primary field of dimension $6/5$
(this is the highest dimension relevant operator in that minimal model). After analyzing the spectra of several candidate minimal models, we suggest that the end point of this RG flow is described by the $(3,8)$ minimal
model with  $c=-21/4$.\footnote{Note that this value is greater than the central charge of the UV theory $M(3,10)$, which is equal to $-44/5$. For flows between non-unitary
theories the Zamolodchikov $c$-theorem does not hold, and it is possible that $c_{UV} < c_{IR}$.}
Let us note that $M(2,5)$ and $M(3,8)$ are members of the series of non-unitary minimal models $M(k,3k-1)$.

In addition to the identity operator, the
$M(3,8)$ model has three Virasoro primary fields which are $Z_2$ odd and three that are $Z_2$ even.
Comparing with the theory in $6-\eps$ dimensions, we can tentatively identify the leading $Z_2$ odd operator as $\phi$ and the leading
$Z_2$ even one as $\sigma$.
Obviously, further work is needed to check if the stable fixed point
in $6-\epsilon$ dimensions with
$g_2^*= 6 g_1^*/5 + O(\eps)$ continued to $\epsilon=4$ is described by the non-unitary minimal model $M(3,8)$.

\section*{Acknowledgments}

We thank J. Gracey and I. Herbut for very useful correspondence, and to Y. Nakayama for valuable discussions.
The work of LF and SG was supported in part by the US NSF under Grant No.~PHY-1318681.
The work of IRK and GT was supported in part by the US NSF under Grant No.~PHY-1314198.

\appendix

\section{Summary of three-loop results}
\label{loopdiagrams}

\begin{figure}[h!]
                \centering
                \includegraphics[width=9cm]{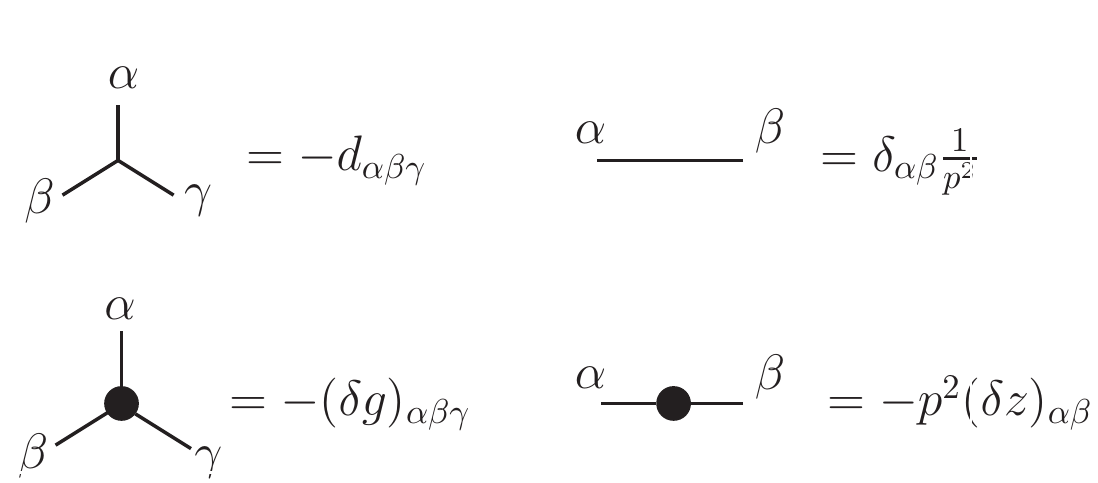}
                 \caption{Feynman rules.}
		\label{Feyrules}
\end{figure}

The Feynman rules for our theory are depicted in Fig.~\ref{Feyrules}

\noindent where we introduced symmetric tensor coupling $d_{\alpha\beta\gamma}$  and counterterms  $(\delta g)_{\alpha\beta\gamma}$, $(\delta z)_{\alpha\beta}$ with $\alpha, \beta, \gamma =0,1,..,N$  as
\begin{align}
&d_{000}=g_{2}, \quad d_{ii0}=d_{i0i}=d_{0ii}=g_{1},  \notag\\
&(\delta g)_{000}=\delta g_{2}, \quad (\delta g)_{ii0}=(\delta g)_{i0i}=(\delta g)_{0ii}=\delta g_{1}, \notag\\
&(\delta z)_{00}=\delta_{\sigma}, \quad (\delta z)_{ii}= \delta_{\phi},
\end{align}
 where $i=1,...,N$.  The general form of a Feynman diagram in our theory could be schematically represented as
\begin{align}
\textrm{Feynman diagram} = \textrm{Integral} \times  \textrm{Tensor structure factor} .
\end{align}
The ``Tensor structure factors'' are  products of the tensors $d_{\alpha\beta\gamma}$ and $(\delta g)_{\alpha\beta\gamma},\, (\delta z)_{\alpha\beta}$,
with summation over the dummy indices. Their values for different diagrams are represented in Fig.~\ref{graph1} and \ref{graph2} after parentheses \footnote{To find the ``Tensor structure factor'' we used the fact that it is  a polynomial in $N$, so we calculated sums of products of $d_{\alpha\beta\gamma}, (\delta g)_{\alpha\beta\gamma}, (\delta z)_{\alpha\beta}$  explicitly  for $N=1,2,3,4,...$, using Wolfram Mathematica.  Having answers for $N=1,2,3,4,..$ it's possible to restore the general $N$ form. }. The ``Integrals'' already include  symmetry factors and are the same as in the usual $\varphi^{3}$-theory; their values are listed in Fig.~\ref{graph1} and \ref{graph2} before the parentheses.

\subsection{Counterterms}
\begin{align}
z_{12}^{\phi} = &-\frac{g_1^2}{3(4\pi)^3}, \quad
z_{12}^{\sigma} = -\frac{Ng_1^2+g_2^2}{6(4\pi)^3},\quad  a_{13}= -\frac{g_1^2 \left(g_1+g_2\right)}{(4\pi)^{3}},\quad
b_{13}= -\frac{N g_1^3 +g_2^3}{(4\pi)^{3}}, \notag\\ \\
z_{14}^{\phi} = &\frac{g_1^2}{432(4\pi)^{6}}  \left(g_1^2 (11 N-26)-48 g_1 g_2 +11 g_2^2\right),   \notag\\
z_{14}^{\sigma} =& -\frac{1}{432(4\pi)^{6}} \left(2N g_1^4 +48 N  g_1^3 g_2-11 N g_1^2 g_2^2+13 g_2^4\right), \notag\\
a_{15}=&-\frac{1}{144(4\pi)^{6}} g_1^2 \big(g_1^3 (11 N+98)-2 g_1^2  g_2  (7 N-38)+101  g_1 g_2^2+4 g_2^3\big), \notag\\
b_{15}=&-\frac{1}{48 (4\pi)^{6}} \big(4N g_1^5 +54 N g_1^4 g_2  +18 Ng_1^3 g_2^2  -7N g_1^2 g_2^3  +23 g_2^5\big), \notag\\ \\
z_{16}^{\phi} =& \frac{g_{1}^{2}}{46656(4\pi)^9}\big( g_1^4(N(13N-232)+5184\zeta(3)-9064) +g_1^3 g_2 6(441N-544)\notag \\
&-2g_1^2 g_2^2(193N-2592\zeta(3)+5881)+ 942g_1 g_2^3+327g_2^4\big), \notag\\
z_{16}^{\sigma}=&-\frac{1}{93312(4\pi)^9}\big(2Ng_1^6(1381N-2592\zeta(3)+4280)-  96N(12N+11)g_1^5 g_2 \notag\\
&-3Ng_1^4 g_2^2(N+4320 \zeta(3)-8882)+ 1560 N g_1^3g_2^3-952N g_1^2 g_2^4 -g_2^6(2592\zeta(3)-5195)\big), \notag\\
a_{17}=&\frac{g_1^2}{15552(4\pi)^{9}}\big(-g_1^5 (N (531 N+10368 \zeta (3)-2600)+23968)\notag\\
&+g_1^4 g_2  (99 N^2+2592 (5 N-6) \zeta (3)-9422 N-2588)+2  g_1^3 g_2^2  (1075 N+2592 \zeta (3)-16897)\notag\\
&+2 g_1^2 g_2^3  (125 N-5184 \zeta (3)-3917)-  g_1 g_2^4 (5184 \zeta (3)+721)+g_2^5 (2592 \zeta (3)-2801)\big), \notag\\
b_{17}=&-\frac{1}{2592(4\pi)^{9}}\big(2 g_1^7 N (577 N+713)-48 g_1^6 g_2  N (31 N-59)+g_1^5 g_2^2  N (423 N+2592 \zeta (3)+1010)\notag\\
&- g_1^4 g_2^3 N (33 N-1296 \zeta (3)-6439)-27 g_1^3 g_2^4  N (32 \zeta (3)+11)-301N g_1^2 g_2^5  +g_2^7 (432 \zeta (3)+1595)\big).
\end{align}

\section{Sample diagram calculations}
\subsection{Some useful integrals}
Many of the diagrams listed in figure \ref{graph1} are recursively primitive, so they can be easily evaluated using the integral:
\begin{equation}
I(\alpha,\beta) = \int \frac{d^{d}p}{(2\pi)^{d}}\frac{1}{p^{2\alpha}(p-k)^{2\beta}} = \frac{L_{d}(\alpha,\beta)}{(k^{2})^{\alpha+\beta-d/2}}\,,
\end{equation}
where
\begin{equation}
L_{d}(\alpha,\beta) = \frac{1}{(4\pi)^{d/2}} \frac{\Gamma(\frac{d}{2}-\alpha)\Gamma(\frac{d}{2}-\beta)\Gamma(\alpha+\beta-\frac{d}{2})}{\Gamma(\alpha)\Gamma(\beta)\Gamma(d-\alpha-\beta)} \,.
\end{equation}

For the more complicated integrals, we use the mathematica program FIRE \cite{Smirnov:2008iw}, which uses integration-by-parts (IBP) relations to turn them into simpler ``master integrals'', which we then evaluate by hand.

\begin{figure}[t]
\centering
\includegraphics[width=12cm]{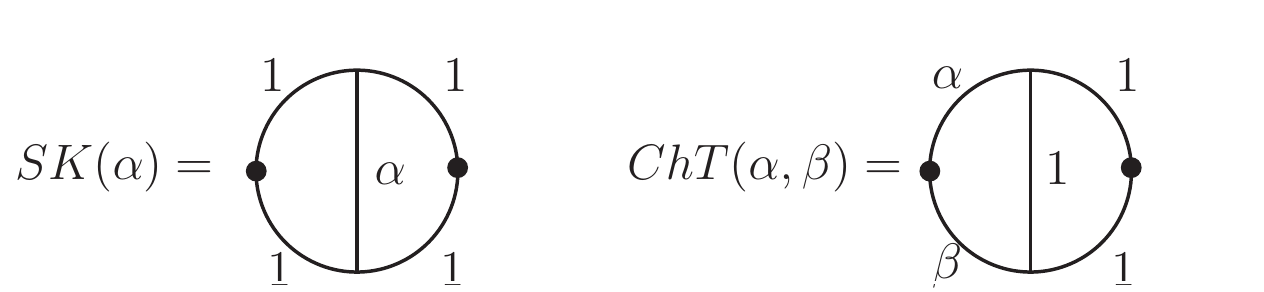}
\caption{The Special $KITE$ and $ChT$ diagrams, the numbers labeling each propagator denote its index.}
\label{SKChT}
\end{figure}

There are two categories of diagrams which show up quite frequently as subdiagrams, the ``special $KITE$'' diagrams and the ``ChT'' diagrams shown in Figure \ref{SKChT}.

The special $KITE$ diagram is a two-loop diagram corresponding to the following integral:
\begin{equation}
SK(\alpha)=\int \frac{d^d p}{(2\pi)^d} \frac{d^d q}{(2\pi)^d} \frac{1}{p^2(p+k)^2q^2(q+k)^2(p-q)^{2\alpha}}\,.
\end{equation}
Notice that the power of the middle propagator is arbitrary. Via the Gegenbauer Polynomial technique as described in \cite{Kotikov:2000yd}, this integral can be expressed as an infinite sum of gamma functions.
\begin{align}
SK(\alpha) &= - \frac{2}{(4\pi)^d} \frac{1}{(k^2)^{4+\alpha-d}}  \frac{\Gamma^2(\lambda)\Gamma(\lambda-\alpha)\Gamma(\alpha + 1 - 2\gamma)}{\Gamma(2\lambda)\Gamma(3\lambda-\alpha-1)}\notag \\
&\times \left(\frac{\Gamma^2(1/2)\Gamma(3\lambda-\alpha-1)\Gamma(2\lambda-\alpha)\Gamma(\alpha+1-2\lambda)}{\Gamma(\lambda)\Gamma(2\lambda+1/2-\alpha)\Gamma(1/2-2\lambda+\alpha)}+\sum\limits_{n=0}^{\infty} \frac{\Gamma(n+2\lambda)}{\Gamma(n+\alpha+1)}\frac{1}{(n+1-\lambda+\alpha)}  \right) \,,
\end{align}
where $\lambda = d/2-1$. In the case of $d=6-\epsilon$ we have found that, for example:
\begin{equation}
SK(2-\frac{d}{2})=\frac{1}{(4\pi)^d}\frac{1}{(k^2)^{6-3d/2}}\bigg( \frac{-54}{\eps^2} + \frac{-71+24 \gamma}{1296\eps} + \frac{-14641+8520 \gamma - 1440 \gamma^2+120\pi^2}{15520}+\ldots \bigg)
\end{equation}
The $\eps$-expansion of the above result can also be verified indirectly with the mathematica packages MBTools implementing the Mellin-Barnes representation \cite{Czakon:2005rk}.

The $ChT$ diagram is another variation of the $KITE$ diagram. It correspond to the integral:
\begin{equation}
ChT(\alpha,\beta)=\int \frac{d^d p}{(2\pi)^d} \frac{d^d q}{(2\pi)^d} \frac{1}{p^{2\alpha}(p+k)^{2\beta}q^2(q+k)^2(p-q)^2}\,.
\end{equation}
In this diagram, one triangle of the $KITE$ diagram all have indices 1, and the other two lines have arbitrary indices $\alpha$ and $\beta$. This diagram was evaluated in position space by Vasiliev et. al. in \cite{Vasiliev:1981dg}.
Their answer is:
\begin{align}
ChT(\alpha,\beta)=&\frac{\pi^d v(d-2)}{\Gamma(\frac{d}{2}-1)} \frac{1}{(x^2)^{d/2-3+\alpha+\beta}} \notag \\
&\times \left(\frac{v(\alpha)v(2-\alpha)}{(1-\beta)(\alpha+\beta-2)}+\frac{v(\beta)v(2-\beta)}{(1-\alpha)(\alpha+\beta-2)}+\frac{v(\alpha+\beta-1)v(3-\alpha-\beta)}{(\alpha-1)(\beta-1)} \right) \,,
\end{align}
where $v(\alpha)=\frac{\Gamma(d/2-\alpha)}{\Gamma(\alpha)}$.
For our purpose, we just need to fourier transform this expression to momentum space.

We also need variations of the $SK$ and $ChT$ diagrams, with a particular index raised by $1$, for example. However, we can use FIRE to relate them to the original version of these diagrams.

\subsection{Example of a two point function diagram}
\begin{figure}[t]
\centering
\includegraphics[width=15cm]{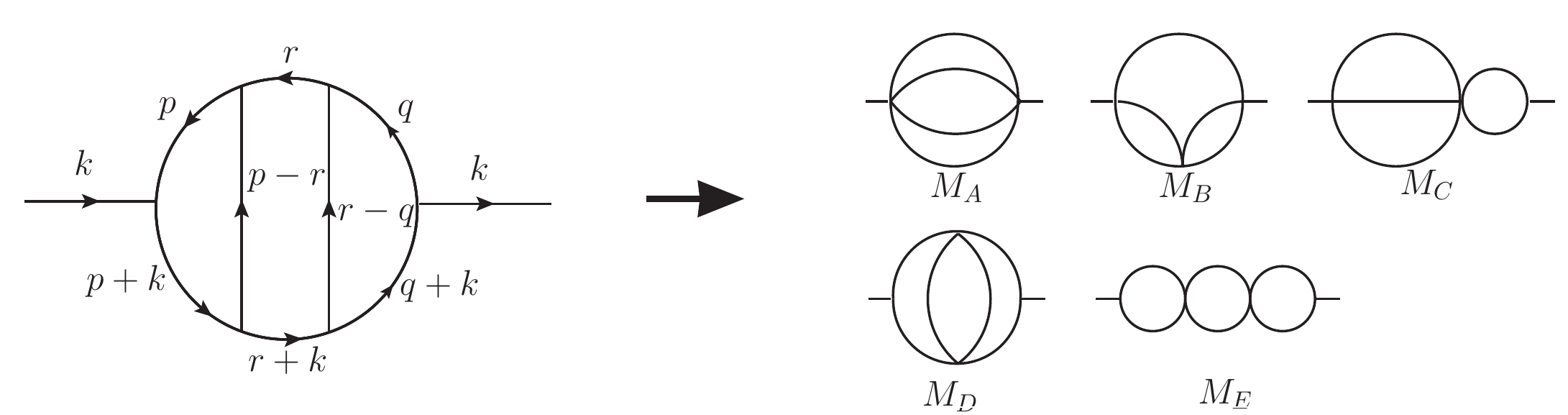}
\caption{The $LADDER$ diagram can be reduced to five master integrals}
\label{LADDER}
\end{figure}

We will evaluate the three-loop ladder diagram which is the first diagram in Figure \ref{graph1}(e). It corresponds to the integral:
\begin{equation}
LADDER = \int \frac{d^dp d^d q d^d r}{(2\pi)^{3d}} \frac{1}{p^2(p+k)^2(p-r)^2r^2(r+k)^2(r-q)^2q^2(q+k)^2}\,,
\end{equation}
where the loop momenta are $p$, $q$, and $r$. The external momentum is $k$.
Using FIRE, it can be reduced to a sum of five master integrals, denoted as $M_A$, ... , $M_E$:
\begin{align}
LADDER =& \frac{4(2d-5)(3d-8)(9d^2-65d+118)M_A}{(d-4)^4k^8}-\frac{12(d-3)(3d-10)(3d-8)M_C}{(d-4)^3k^6} \notag \\
&+ \frac{32(d-3)^2(2d-7)M_B}{(d-4)^3k^6} + \frac{4(d-3)^2M_E}{(d-4)^2k^4}+\frac{3(d-3)(3d-10)M_D}{(d-4)^2k^4}\,.
\end{align}
The diagrams corresponding to the master integrals are listed in Figure \ref{LADDER}.
Among these master integrals, only $M_D$ is non-primitive, the rest can be calculated easily. However, if we integrate over the middle loop, we see that $M_D$ is in fact related to the special $KITE$ diagram $SK(2-d/2)$. We have:
\begin{align}
M_A &= \frac{L_{d}(1,1)L_{d}(1,2-d/2)L_{d}(1,3-d)}{(k^2)^{4-3d/2}}, \quad
M_B = \frac{(L_{d}(1,1))^{2}L_{d}(1,4-d)}{(k^2)^{5-3d/2}} \notag \\
M_C &= \frac{(L_{d}(1,1))^{2}L_{d}(1,2-d/2)}{(k^2)^{5-3d/2}},\\
M_D &= L_{d}(1,1) SK(2-d/2), \quad
M_E = \frac{(L_{d}(1,1))^3}{(k^2)^{6-3d/2}} \notag.
\end{align}
Plugging in $d=6-\eps$ and expanding in $\eps$, we find that:
\begin{align}
LADDER =& \frac{k^2}{(4\pi)^{3d/2}} \bigg( -\frac{2}{9\eps^3} + \frac{-115 + 36\gamma + \log{k^2}}{108 \eps^2} \notag \\
&+ \frac{-4043+18(115-18\gamma)\gamma + 18\pi^2-18\log{k^2}(-115 + 36 \gamma + 18\log{k^2}) }{1296 \eps} + \ldots \bigg)\,.
\end{align}

\subsection{Example of a three point function diagram}
We will evaluate the three-loop diagram found in Figure \ref{graph2}(f).
In order to employ the same techniques used for the two-point functions, we impose that the momentum running through the three points are $p$, $-p$, and $0$, as a three-point function with three arbitrary momenta are much more difficult to compute.

\begin{figure}[t]
\centering
\includegraphics[width=10cm]{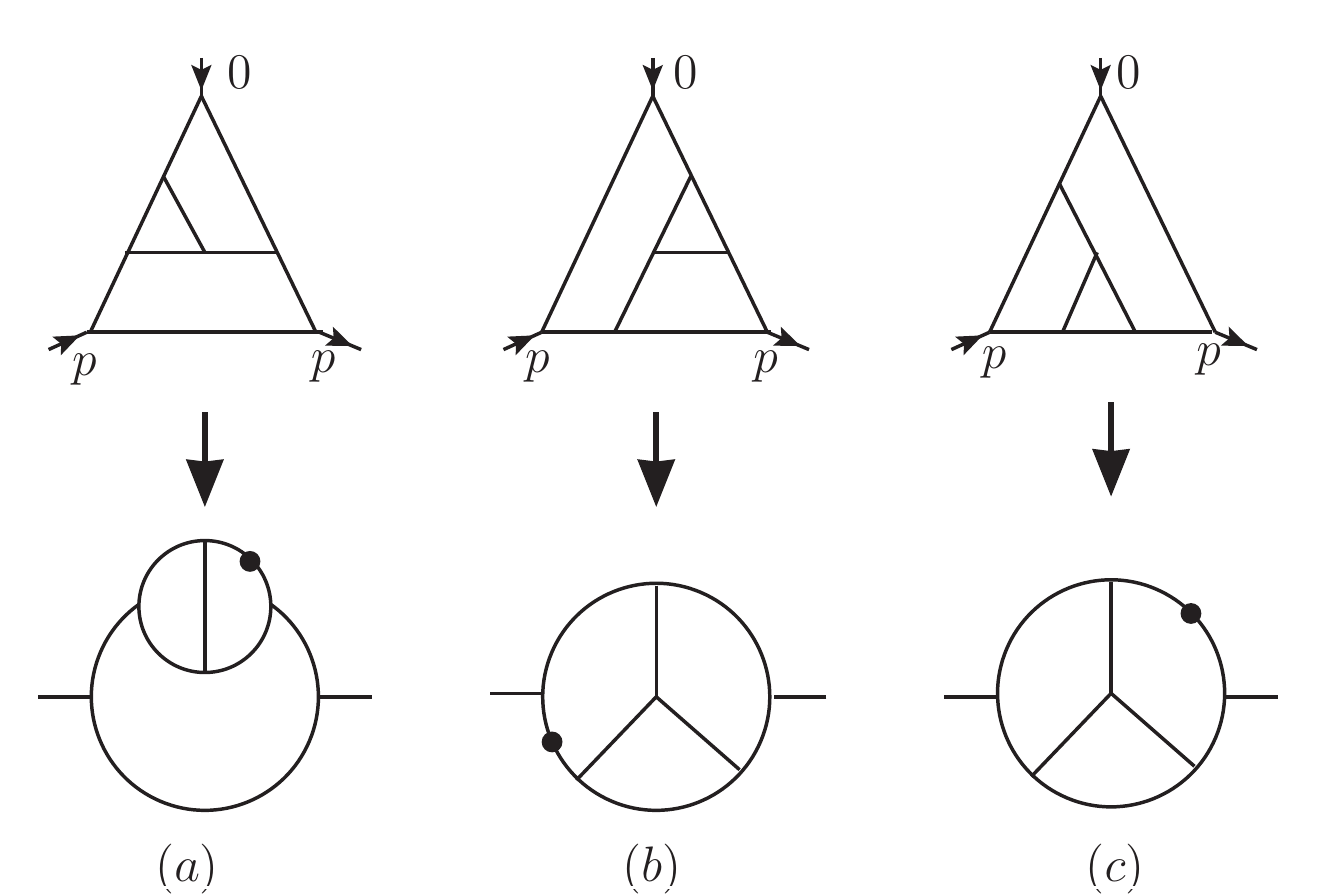}
\caption{The three orientations of the same diagram topology correspond to different integrals.}
\label{Int17}
\end{figure}

However, since the momenta are asymmetric, it is necessary to consider all three ``orientations'' of each topology of the diagram. Notice that the tensor factors mentioned in the previous section will also be different. As an illustration, let's denote the three orientations of the diagram we are considering by $I_1$, $I_2$, and $I_3$. After taking into account that one of the external momenta is zero, they are each equivalent to a two-point function as shown in Figure \ref{Int17}. All lines have indices 1, except those with black dots, which have indices 2.

$I_1$ contains a subdiagram that is equivalent to $ChT(1,2)$, which can be evaluated easily using our formula before; after that, the diagram is primitive. The other two diagrams be reduced via FIRE into the master integrals $M_A$, $M_B$, $M_C$, and $M_D$ as in the $LADDER$ diagram.
Again, in $d=6-\eps$, we find that:
\begin{align}
I_1 =& \frac{1}{(4\pi)^{3d/2}} \bigg( \frac{1}{6 \eps^3} + \frac{5-2\gamma-2\log{k^2}}{8\eps^2} \notag\\
&+ \frac{173+18(\gamma-5)\gamma - \pi^2 + 18\log{k^2}(-5 +2\gamma+\log{k^2})}{96\eps} + \ldots \bigg) \\
I_2 =& \frac{1}{(4\pi)^{3d/2}} \bigg( \frac{1}{6 \eps^3} + \frac{5-2\gamma-2\log{k^2}}{8\eps^2} \notag\\
&+ \frac{125+18(\gamma-5)\gamma - \pi^2 + 18\log{k^2}(-5 +2\gamma+\log{k^2})}{96\eps} + \ldots \bigg) \\
I_3 =& \frac{1}{(4\pi)^{3d/2}} \bigg( \frac{1}{6 \eps^3} + \frac{5-2\gamma-2\log{k^2}}{8\eps^2} \notag\\
&+ \frac{125+18(\gamma-5)\gamma - \pi^2 + 18\log{k^2}(-5 +2\gamma+\log{k^2})}{96\eps} + \ldots \bigg)\,.
\end{align}

 \begin{landscape}

\begin{figure}[h!]

                \centering
                \includegraphics[width=25cm]{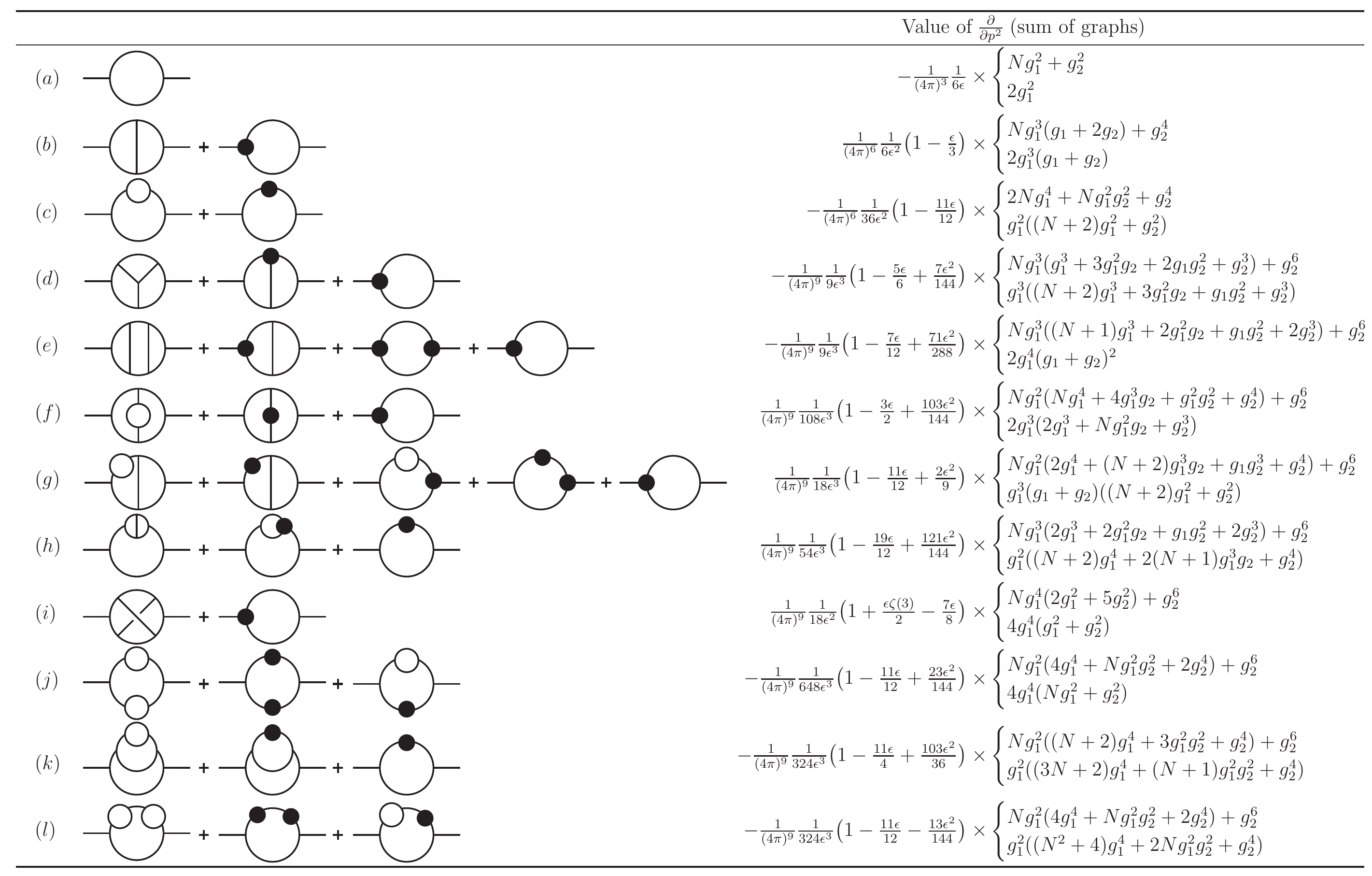}
                 \caption{Values of derivatives of two-point diagrams. The upper row in parenthesis  is for $\langle \sigma \sigma \rangle $  and the lower is for $\langle \phi \phi \rangle$.}
                 \label{graph1}
\end{figure}

\begin{figure}[h!]
               \caption{Values of three-point diagrams. The upper row in parenthesis  is for $\langle \sigma \sigma \sigma \rangle $  and the lower is for $\langle \sigma \phi \phi \rangle$.}
                \centering
                \includegraphics[width=25cm]{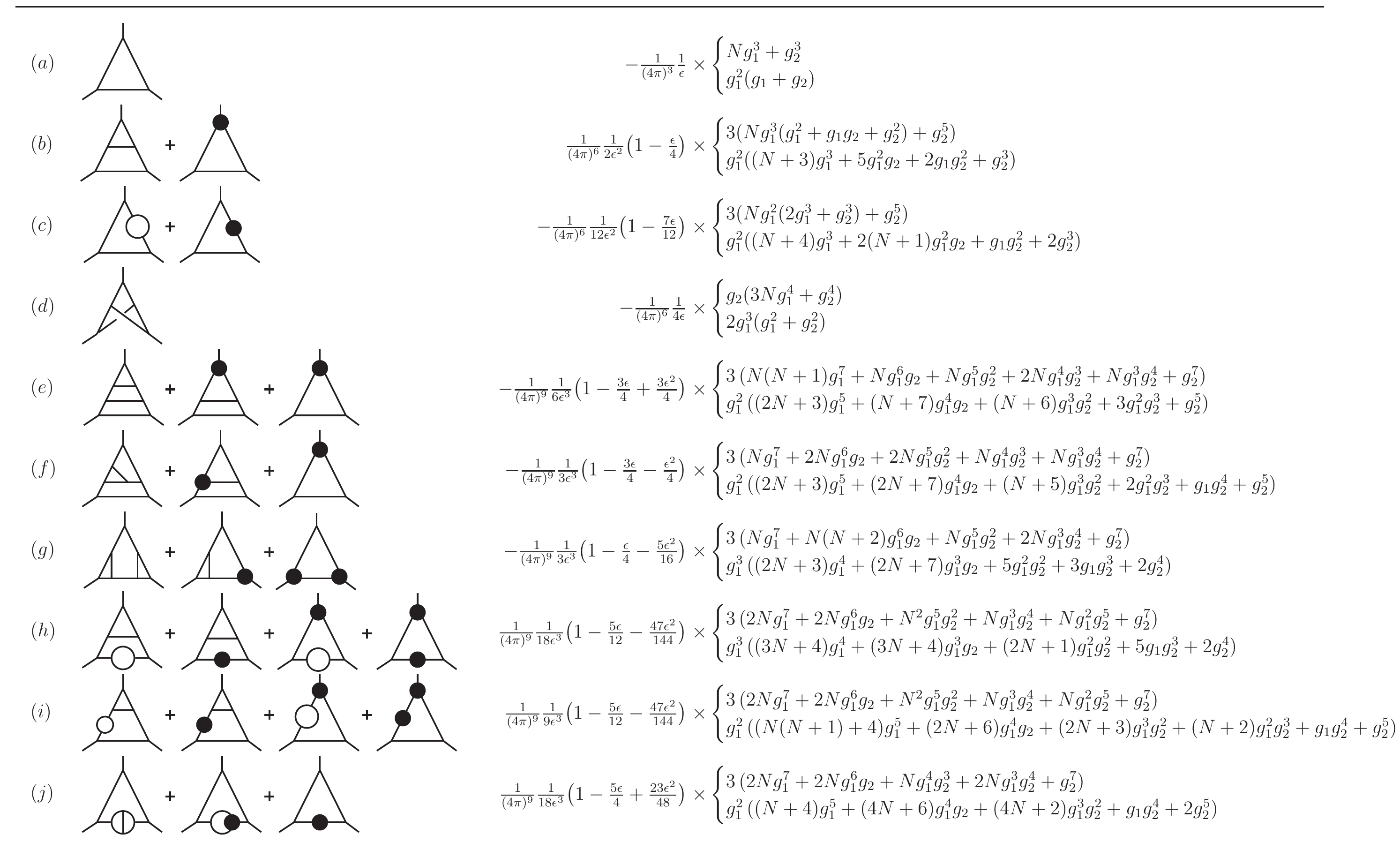}
                                \label{graph2}
\end{figure}

\begin{figure}[h!]
                \centering
                \includegraphics[width=25cm]{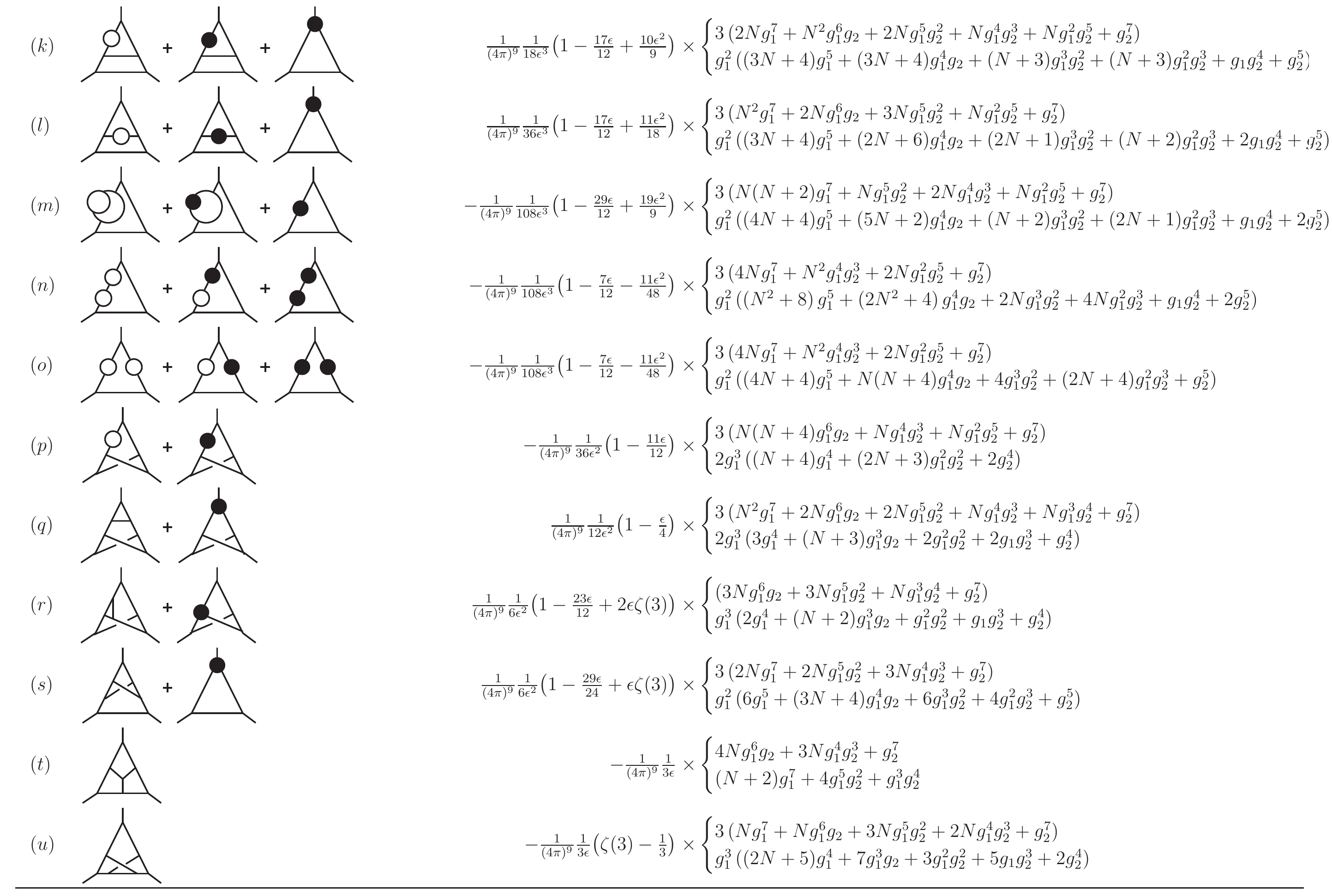}
                           \caption{}

\end{figure}

\end{landscape}


\bibliographystyle{ssg}
\bibliography{beta_3loop}

\end{document}